\documentclass[traditabstract]{aa}

\usepackage{graphicx}
\usepackage{longtable}

\usepackage{hyperref}
\usepackage{natbib}                 
\bibpunct{(}{)}{;}{a}{}{,}          
\usepackage{amsmath}                

\usepackage{txfonts}
\begin{document}

   \title{1ES 0229+200: An extreme blazar with a very high minimum Lorentz factor}

 {\small  \author{
S.~Kaufmann\inst{1}
\and S.J.~Wagner\inst{1}
 \and O.~Tibolla \inst{2}
\and M.~Hauser \inst{1}
}
}

   \institute{
Landessternwarte, Universit\"at Heidelberg, K\"onigstuhl, D-69117 Heidelberg, Germany
\and
Universit\"at W\"urzburg, 97074 W\"urzburg, Germany
}


   \date{Received 9 May 2011; accepted 18 August 2011}


  \abstract
%
{The blazar 1ES 0229+200 is a high frequency peaked BL Lac object with a hard TeV spectrum extending to $10 \; \rm{TeV}$. Its unusual spectral characteristics make it a frequently used probe for intergalactic radiation and magnetic fields. 
With new, simultaneous observations in the optical, ultraviolet (UV) and X-rays, the synchrotron emission is probed in great detail. 
The X-ray emission varies by a factor of $\approx 2$ in 2009, while being rather stable in 2010. 
The X-ray spectrum is very hard ($\Gamma \approx 1.8$) and it shows an indication of excess absorption above the Galactic value.
The X-ray emission is detected up to $\sim 100\;\rm{keV}$ without any significant cut-off, thus 1ES 0229+200 belongs to the class of extreme blazars. 
The simultaneous measured, host galaxy- and extinction-corrected optical and UV fluxes illustrate that the cut-off of the low energy part of the synchrotron emission is located in the UV regime. The minimum energy of the electron distribution has to be rather high to account for this cut-off. 
This implies that there is a narrow-band energy distribution function of radiating electrons, which is responsible for the unusually hard TeV spectrum.
Other extreme blazars have similar synchrotron peak frequencies but much softer TeV spectra, hence 1ES 0229+200 has one of the highest inverse Compton (IC) peak frequency and the narrowest electron distribution among the extreme blazars known to date.
}
   \keywords{Galaxies: active - BL Lacertae objects: Individual: 1ES0229+200 - X-rays, UV, optical: observations}

   \authorrunning{S. Kaufmann et al.}

   \maketitle
%

\section{Introduction}

The high-frequency peaked BL Lac object 1ES 0229+200 is located at $\alpha_{\rm{J2000}} = 2^{\rm{h}} 32^{\rm{m}} 48.62^{\rm{s}}$, $\delta_{\rm{J2000}} = +20^\circ 17' 17.45''$ \citep{Rector2003} and has a redshift of $z=0.14$ \citep{Woo2005}.

High frequency peaked BL Lac objects (HBL) are characterized by two peaks in their spectral energy distribution (SED) which are located in the UV-X-ray and the GeV-TeV band, respectively. 
These are commonly interpreted in terms of leptonic models (e.g. \citealt{Marscher1985}) as synchrotron and inverse Compton (IC) emission from a population of relativistic electrons upscattering their self-produced synchrotron photons (synchrotron self-Compton (SSC) models).

1ES 0229+200 was discovered in the  {\it Einstein} IPC Slew Survey \citep{Elvis1992}, and classified as a high-frequency peaked BL Lac object based on its X-ray to radio flux ratio \citep{Giommi1995}.

The VLA observations of 1ES 0229+200 reveal a core flux of $51.8\;\rm{mJy}$ and show curved jets with an extension of $\sim 30''$ at 1.4 GHz and jet position angles of P.A.$=-10^\circ$ and P.A.$=180^\circ$ \citep{Rector2003}. 

1ES 0229+200 is not detected in the high energy $\gamma$-ray range (100 MeV
$<$ E $<$ 100 GeV) by {\it Fermi}/LAT in two years of observations and hence it is not mentioned in the second {\it Fermi} catalog \citep{Abdo2011}.

In 1996, it was originally predicted to be a potential VHE $\gamma$-ray source based on its SED \citep{Stecker1996}, however, Whipple, HEGRA, and Milagro have only reported upper limits (\cite{Horan2004}, \cite{Aharonian2004a},\cite{Williams2005}).
Very high-energy (VHE, E$>$100 GeV) emission up to $10\; \rm{TeV}$ was first detected with the High Energy Stereoscopic System (H.E.S.S.) in 2006  \citep{Aharonian2007}. In this study, a hard spectrum with a photon index of $\Gamma = 2.5 \pm 0.19_{\rm{stat}} \pm 0.1_{\rm{sys}}$ was reported.
Besides 1ES~1426+428 (\cite{Aharonian2003}), 1ES 0229+200 is the only source at redshift $z>0.1$
whose spectrum was measured up to this high energy.
1ES 0229+200 has a very hard intrinsic VHE spectrum,
hence it is well-suited to EBL studies (e.g. \cite{Aharonian2007}, \cite{Kneiske2010}).
Its spectral characteristics have been used to probe intergalactic magnetic fields (e.g. \cite{Neronov2010}, \cite{Tavecchio2010}).
At the same time, these hard spectra are challenging for blazar models. We have
launched dedicated, simultaneous multi-wavelength observations with XMM-Newton (X-ray, UV) and ATOM (optical) to determine the broad-band spectra in more detail.

\section{Multi-wavelength observations and data analysis}

\begin{table*}
\centering
\begin{tabular}{c|c|c|c|c|c|}
Instrument       & Time & ObsID & Energy range & Photon index & $F_{\rm{2-10 keV,deabs}}\; (\rm{erg \; cm^{-2} \; s^{-1}})$\\      
\hline
{\it XMM-Newton}        & 21.,23.August 2009 & 604210201,604210301 & 0.1-15 keV & $1.84 \pm 0.02$ & $(9.0 \pm 0.1)\times10^{-12}$ \\
{\it Swift} /XRT       & 5.,7.,8. August 2008 & 31249001-31249003 & 0.2-10 keV & $1.87 \pm 0.05$ & $(1.03 \pm 0.04)\times10^{-11}$ \\
{\it Swift} /XRT       & 19.Oct. - 23. Nov. 2009 & 31249004-31249019 & 0.2-10 keV & $1.73 \pm 0.03$ & $(1.47 \pm 0.04)\times10^{-11}$ \\
{\it RXTE} /PCA       & 1.Jan. - 13. Oct. 2010 & 95387 & 3-60 keV & $1.92 \pm 0.05$  & $(1.23 \pm 0.04)\times10^{-11}$  \\
{\it BeppoSAX}       & 16. July 2001 & 51472001 & 0.1-50 keV & $1.99 \pm 0.05$ & $1.5\times10^{-11}$\\
{\it ROSAT}       &  &  & 0.1-2.4 keV &  & $(4.5 \pm 0.6)\times10^{-12}$ at 1keV \\
{\it Einstein}        &  &  & 0.4-4.0 keV &   & $(7.6 \pm 2.2)\times10^{-12}$ at 1keV \\
\hline
\end{tabular}
\caption{X-ray observations of 1ES 0229+200 showing the results for the annual binning of {\it XMM-Newton}, {\it Swift}, and {\it RXTE}, as well as historical observations from {\it Einstein} \citep{Elvis1992}, {\it ROSAT} \citep{Brinkmann1995} and {\it BeppoSAX} \citep{Donato2005}. 
The measured fluxes are given for the energy range between 2 and 10 keV. For the {\it ROSAT} and {\it Einstein} observations, the fluxes are determined at an energy of 1keV.
The absorption derived from the high precision {\it XMM-Newton}  spectra of $N_{\rm{H}} = 1.1\times 10^{21} \; \rm{cm^{2}}$ was used for the {\it XMM-Newton}, {\it Swift},  and {\it RXTE}  spectra to determine the unabsorbed flux. The absorption detected in the {\it BeppoSAX} spectrum was $N_{\rm{H}} = 1.0\times 10^{21} \; \rm{cm^{2}}$ \citep{Donato2005}.
}
\label{tab_obs}
\end{table*}

\subsection{X-ray data from {\it XMM-Newton}, {\it Swift},  and {\it RXTE} }

{\it XMM-Newton}  observations of 1ES 0229+200 were carried out on August 21 and 23, 2009 for 23 and 28 ks, respectively. The observations were conducted with MOS1 and PN in full imaging mode and MOS2 in timing mode, all with a thin filter. The two grating spectrometers onboard {\it XMM-Newton}  RGS 1, 2 were also used to acquire. 
The data analysis of the {\it XMM-Newton}  observations was performed with SAS v.9.0.
The data from the X-ray instruments were reprocessed as described in the SAS user-guide\footnote{http://xmm.esac.esa.int/external/xmm\_user\_support/documentation\\/sas\_usg/USG/}.
Both observations are influenced by short soft proton flares at the end of each observation, detected in the high $>12\;\rm{keV}$ emission for the PN ($>10\;\rm{keV}$ for MOS) detector.
Therefore, the good time intervals were determined using a cut at 0.4 cts/s for PN (0.35 cts/s for MOS) for the high-energy count rate resulting in an effective exposure of 7.3 ks for PN (23.1 ks for MOS) for the first and 12.1 ks for PN (24.4 ks for MOS) for the second pointing.
No significant variations were found in either observation using different binning down to a timescale of 100 (10) seconds for the imaging mode (timing mode) during these pointings. 

For the imaging mode, spectra were extracted  from a source region of $30''$ radius for the PN ($80''$ for MOS) around the position of 1ES 0229+200 and background regions with the same radii, 
on the same chip of the detectors that contained the source region.
No significant pileup was found in the spectra of the source regions using the tool {\tt epatplot}.
The RMF (redistribution matrix file) and ARF (ancillary response file) are calculated for each spectrum using the tools {\tt rmfgen} and {\tt arfgen}. 


The grating spectrometers RGS 1, 2 onboard {\it XMM-Newton}  measure in the energy range 0.35 to 2.5 keV. The data are analysed using the tool {\tt rgsproc}. 

From October 19 to November 23, 2009, 16 {\it Swift}  observations were targeting on 1ES 0229+200.
During these pointings with observations of $0.5-4\;\rm{ks}$ each, the XRT detector \citep{Burrows2005} was operated in windowed-timing (WT) mode in the energy range $0.2-10\;\rm{keV}$. 
In 2008, 1ES 0229+200 was observed  on August 5, 7, and 8 in photon-counting
(PC) and WT mode. However, only the PC mode observations are taken into
account in the analysis, since only 120 s were observed in WT mode resulting
in an insufficient amount of counts.
For the {\it Swift}  spectral analysis, XRT exposure maps were generated with the {\tt xrtpipeline} to account for some bad CCD columns that are masked out on-board. The masked hot columns appeared when the XRT CCD was hit by a micrometeoroid.
Spectra of the {\it Swift} data in PC-mode have been extracted with {\tt xselect} from an annulus region with a radius of $0.8'$ at the position of 1ES 0229+200, which contains $90\%$ of the PSF at 1.5~keV. The background was extracted from a circular region  with radius of $3'$ near the source. For the WT-mode, boxes ($\sim 1.6' \times 0.3'$) covering the region with source photons and a background region of similar size were used to extract the spectra. 
The auxiliary response files were created with {\tt xrtmkarf} and the response matrices were taken from the {\it {\it Swift} } package of the calibration database {\tt caldb 4.1.3}\footnote{http://heasarc.gsfc.nasa.gov/docs/heasarc/caldb/caldb\_intro.html}.

X-ray observations with the Proportional Counter Array (PCA) detector onboard {\it {\it RXTE} } \citep{Bradt1993} were obtained in the energy range $2-60\;\rm{keV}$ from January 1 to October 13 2010 
with exposures of $1-2\;\rm{ks}$ per pointing.
Only {\it {\it RXTE} }/PCA data of PCU2 and the top layer 1 were considered to ensure the highest signal-to-noise ratio. The data were filtered to account for the influence of the South Atlantic Anomaly, tracking offsets, and electron contamination using the standard criteria recommended by the {\it {\it RXTE} } Guest Observer Facility (GOF). For the count rate of $\sim 1\;\rm{cts/s}$ for this observations, the faint background model, provided by the {\it {\it RXTE} } GOF was used to generate the background spectrum with {\tt pcabackest} and the response matrices were created with {\tt pcarsp}.

The second instrument onboard {\it RXTE}, the HEXTE (High Energy X-ray Timing Experiment) onboard {\it RXTE} takes data in the energy range 15 to 250 keV. 
Since 2006, the HEXTE cluster A has been operating only in ON-source mode and since the end of 2009 the cluster B 
is permanently operating in OFF-source mode. 
For the spectral analysis, the cluster B data are used as background information for the cluster A data. The sum of all observations from January 1 until October 13, 2010 show no significant signal from 1ES 0229+200 in this energy range.

\subsection{Additional X-ray sources in the {\it XMM-Newton} field of view}

The source detection tool {\tt edetect\_chain} revealed 
20 point sources 
in the field of view of the PN detector and one source that is extended beyond the PSF of the instrument. 
Two point sources are coincident with IRAS sources, four point sources are coincident with NVSS sources, and 15 with stars from the Guide Star Catalogue (GSC, \cite{Lasker2008}) (see table in online supplement).
The remaining point sources do not have a counterpart in these catalogues and remain unidentified.

The extended source XMMU~023318.0+201237 is located at  $\alpha_{\rm{J2000}} = 2^{\rm{h}} 33^{\rm{m}} 18.05^{\rm{s}}$, $\delta_{\rm{J2000}} = +20^\circ 12' 37.19''$ and is very close ($ 58''$ offset) and likely connected to a point-like source  XMMU 023315.5+201323 that is positional coincident with an infrared source IRAS 02304+2000 with coordinates $\alpha_{\rm{J2000}} = 2^{\rm{h}} 33^{\rm{m}} 14.6^{\rm{s}}$, $\delta_{\rm{J2000}} = +20^\circ 13' 30''$ and a flux of 0.4 Jy at 60 $\mu m$ \citep{Moshir1990}. This infrared (IR) source has a radio counterpart NVSS023314+201330 that has an elliptical shape in the NVSS sky map and a flux of $12.4\pm1.1$ Jy at 1.4 GHz \citep{Condon1998}.
One of the R-band observations by ATOM cover the region of this source and an object with 16.5 mag is measured. A closer look yields the detection of two very close ($4''$ separation) point sources of which the brighter one can be identified with USNO-B1.0 1102-0028956.

\subsection{UV data from {\it XMM-Newton}/OM and {\it Swift}/UVOT}

The optical monitor (OM) \citep{Mason2001} onboard {\it {\it XMM-Newton} } observed 1ES 0229+200 in the filters UVM2 (231 nm), UVW1 (291 nm), and U (344 nm) simultaneously with the X-ray telescope. The analysis of these data was performed with  the analysis described in the webpage\footnote{http://xmm.esac.esa.int/sas/current/documentation/\\threads/omi\_stepbystep.shtml}.

The UVOT instrument \citep{Roming2005} onboard {\it {\it Swift} } measures the UV and optical emission in the bands UVW2 (188 nm), UVM2 (217 nm), UVW1 (251 nm), U (345 nm), B (439 nm), and V (544 nm) simultaneously with the X-ray telescope with an exposure of $0.4-3\;\rm{ks}$ each. The instrumental magnitudes and the corresponding flux (see \citealt{Poole2008} for the conversion factors) are calculated with {\tt uvotmaghist} taking into account all photons from a circular region of radius $5''$ (standard aperture for all filters). An appropriate background was determined from a circular region of radius $5''$ near the source region without contamination of sources.

\subsection{Optical data from ATOM}

The 75-cm telescope ATOM \citep{Hauser2004}, located at the H.E.S.S. site in Namibia, monitored the flux in the different filters B (440 nm), V (550 nm), R (640 nm), and I (790 nm) according to \cite{Bessel1990}. The obtained data were analysed using an aperture of $4''$ radius. Photometric calibration was done using the standard fields SA\,113 and SA\,95 from \cite{Landolt1992}.




\section{Spectral data in the synchrotron range}

\begin{figure}[h]
\centering
\includegraphics[width=\columnwidth]{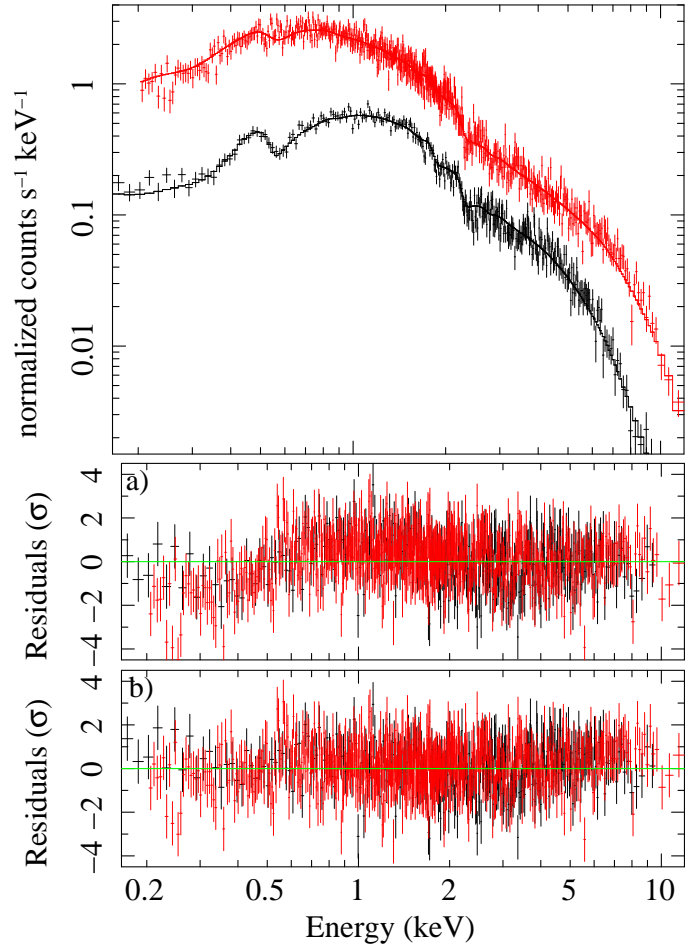}
\caption{{\it XMM-Newton}  MOS1 (black) and PN (red) spectrum of 1ES 0229+200 from August 21, 2009. The spectra can be well fit with a power-law model taking into account an absorption larger than the Galactic absorption. In panel {\it a}, we plot the residuals for a power law considering the Galactic absorption as fixed parameter, and in panel {\it b} residuals for a power law with a free absorption.} 
\label{XMM_spec}
\end{figure}

All {\it XMM-Newton}, {\it Swift}, and {\it RXTE} spectra are binned 
with at least 25 counts and {\tt xspec v12.5} was used for the spectral analysis. 
For the {\it XMM-Newton}  spectra, the energy ranges were restricted to $0.1 -10 \; \rm{keV}$ for MOS and $0.2-15\;\rm{keV}$ for PN following the suggestion of the calibration information\footnote{XMM-SOC-CAL-TN-0018: \\http://xmm.vilspa.esa.es/docs/documents/CAL-TN-0018.pdf}. 
The MOS1 and PN spectra are fitted simultaneously using a constant parameter to account for the different normalizations.
A power-law model of the form $F(E) = N_0(E/E_0)^{-\Gamma}$ was used to fit the X-ray spectra taking into account the Galactic absorption of $N_{\rm H} = 7.9\times 10^{20} \; \rm{cm^{2}}$ (LAB Survey, \cite{Kalberla2005}). As can be seen in Fig. \ref{XMM_spec}a, the residuals of the fit to the X-ray spectra deviate from the expected form, which implies that their is either additional absorption or  different spectral characteristics.

Fitting a power-law model with unconstrained absorption results in 
a much better description of the data with $\chi^2 / dof = 979.7/903 = 1.08$. 
The fit parameter and the goodness of each model is given in table \ref{tab_resuXMM}.
Residuals are shown in panel b of Fig. \ref{XMM_spec}.

Alternatively, the deviations shown in Fig.~\ref{XMM_spec} could be avoided by 
taking into account the Galactic absoption and
fitting a broken power-law. 
This fit results in photon indices of $\Gamma_1 =0.8 \pm 0.3$, $\Gamma_2 = 1.78 \pm 0.01$, and a break at $0.58 \pm 0.09 \; \rm{keV}$  ($\chi^2 / dof = 935/887$). 
However, the low energy extrapolation of the X-ray spectrum would not fit the UV-optical range, hence this model is disfavoured.

To identify the location of the additional absorption, we test the hypotheses that either the total $N_{\rm H}$ is located at redshift z=0 or that the additional absorption is located at redshift z=0.14.
The resulting models with two different redshifts are indistinguishable and have similar goodnesses of fit (at z=0.14: $\chi^2 / dof = 961.3 / 888$; at z=0: $\chi^2 / dof = 959.3 / 888$). Hence the location of the additional absorption could be either intrinsic to 1ES 0229+200 or in the line of sight to the observer or in the Milky Way.
We note that both a local enhancement of Galactic  $N_{\rm H}$ by $25\%$ along the line of sight to 1ES~0229+200, as well as an intrinsic column within the source of $2.9\times 10^{20} \; \rm{cm^{2}}$, are plausible.

\begin{table*}
\centering
\begin{tabular}{|c|c|c|c|c|c|c|}
\hline
Date    &  $N_{\rm{H}} $ & $\Gamma$ & Norm. & Norm. factor & $\chi^2 / dof $ & $F_{2-10 \;\rm{keV}} $ \\      
 & $ (\rm{cm^{2}})$ & & $ (\rm{ph \; cm^{-2} \; s^{-1} \; keV^{-1}})$ & (PN) &  & $ (\rm{erg \; cm^{-2} \; s^{-1}})$ \\
\hline
21. August 2009 & $ 7.9\times 10^{20}$ &$1.72 \pm 0.01$ & $(2.44 \pm 0.03)\times 10^{-3}$ & $0.990$ & $1125.5/889$ & $(9.6 \pm 0.1)\times 10^{-12}$ \\
 & $(1.08 \pm 0.04)\times 10^{21}$ & $1.84 \pm 0.02$ & $(2.73 \pm 0.05)\times 10^{-3}$ & $1.004$ & $959.3/888$ & $(8.9 \pm 0.1)\times 10^{-12}$ \\
\hline
23. August 2009 & $ 7.9\times 10^{20}$ &$1.72 \pm 0.01$ & $(2.49 \pm 0.02)\times 10^{-3}$ & $0.974$ & $1376.9/1064$ & $(9.8 \pm 0.1)\times 10^{-12}$ \\
 & $(1.06 \pm 0.03)\times 10^{21}$ & $1.83 \pm 0.02$ & $(2.77 \pm 0.04)\times 10^{-3}$ & $0.989$ & $1133.7/1063$ & $(9.1 \pm 0.1)\times 10^{-12}$ \\
\hline
\end{tabular}
\caption{{\it XMM-Newton}  fit parameter, goodness of fit, and unabsorbed flux resulting from a simultaneous fit of the MOS1 and PN X-ray spectra. The absorption was fixed to the Galactic absorption of the LAB survey \citep{Kalberla2005} in the first model and let free to vary for the second resulting in additional amount of absorption.
}
\label{tab_resuXMM}
\end{table*}

The {\it XMM-Newton}/RGS spectra show no significant line emission and the continuum spectra are well-described by a power law with additional absorption comparable to the PN and MOS spectra.

The spectrum of the extended source detected in the PN and MOS observations close to 1ES 0229+200, was extracted from a region of radius $1.7'$ for the PN detector ($1'$ for the MOS detector due to the CCD gaps) and reveals $90\%$ and $30\%$a faint source with a flux of $\approx 1\times 10^{-13}\;\rm{erg \; cm^{-2} \; s^{-1}}$. We were able to fit the spectrum with a binning of at least 15 photons per bin with a fixed Galactic absorption and a power law model with $\Gamma =2.0 \pm 0.2$ ($\chi^2/dof = 73/96$) and slightly better by a thermal model ({\tt mekal}) with $kT = 4\pm 2\; \rm{keV}$ ($\chi^2/dof = 87/96$).

\begin{figure}[h]
\centering
\includegraphics[width=\columnwidth]{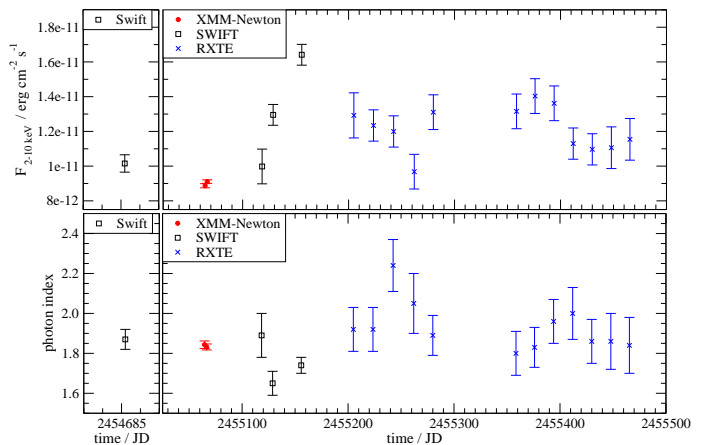}
\caption{The X-ray flux of 1ES 0229+200 varied by a factor of $\sim 2$ in October 2009 as shown in the combined lightcurve taken with the instruments {\it XMM-Newton} (red cirlces), {\it Swift}  (black open squares), and {\it RXTE}  (blue crosses) in the energy range 2-10 keV. For comparison, the {\it Swift}  observation of 2008 is also shown. The lower panel displays the spectral indices derived by fitting power-law models to the spectra.
} 
\label{Xray_LC}
\end{figure}

The annual averages of the {\it Swift} spectra in 2008 and 2009 are shown in table~\ref{tab_obs}. The consecutive pointings of the 2009 data obtained between October and November were binned in three ten-day intervals each to increase the statistics and are shown in Fig.~\ref{Xray_LC}. 
Initially, a simple power law and a free absorption was used to fit these spectra and result in the best description of the spectral shape. 
No significant change in absorption was detected with values of $N_{\rm H} = (1.6 \pm 0.4)\times 10^{21} \; \rm{cm^{2}}$ in 2008 and $N_{\rm H} = (1.3 \pm 0.5)\times 10^{21} \; \rm{cm^{2}}$, $N_{\rm H} = (0.9 \pm 0.5)\times 10^{21} \; \rm{cm^{2}}$ and $N_{\rm H} = (1.3 \pm 0.2)\times 10^{21} \; \rm{cm^{2}}$ for the spectra of 2009 binned in ten-day intervals. These values are comparable to the one obtained from the {\it XMM-Newton} spectra, which provide the most precise determination of the additional absorption. A total column of $N_{\rm H} = 1.1\times 10^{21} \; \rm{cm^{2}}$ was assumed to obtain the photon indices and the fluxes (see table~\ref{tab_obs} and Fig.~\ref{Xray_LC}).
The {\it Swift} data from August 2008 were analysed and are shown in the left panel in Fig.~\ref{Xray_LC} (2454685 JD). It should be noted here that these data were already presented in \cite{Tavecchio2009}. Our re-analysis reveals that the integrated flux between 2 and 10 keV is lower by a factor of $\sim 7$, but confirms the spectral indices. This is independent of the choice of a higher value of $N_{\rm H}$.



The {\it RXTE}  spectra have been summed covering about 30 days to achieve higher statistics. The energy range from 2-3 keV and 20-30 keV were excluded from the spectral analysis, because of instrumental features (e.g. spikes in the background files, see website\footnote{http://www.universe.nasa.gov/xrays/programs/{\it RXTE} /pca/doc/\\bkg/bkg-2009-spikes/}).
A simple power-law was used to fit these spectra. The resulting photon indices and fluxes can be seen in Fig.~\ref{Xray_LC}.


The historical spectrum of {\it BeppoSAX} was fitted with a power law of photon index $\Gamma = 1.99 \pm 0.05$ and a free absorption of $N_{\rm{H}} = (10 \pm 5) \times 10^{20} \;\rm{cm^{-2}}$ in the energy range $0.1 - 50\; \rm{keV}$ without any indication for a  cut-off at the high energy end \citep{Donato2005}.

The {\it Swift}/BAT spectrum taken from the 58 month catalog\footnote{http://heasarc.gsfc.nasa.gov/docs/{\it Swift} /results/bs58mon/} (\cite{Baumgartner2010}) is well fit by a power law with photon index of $\Gamma = 2.1 \pm 0.3$ ($\chi^2_{\rm{red}} = 1.02$) and result in a flux between 14 and 195 keV of $F = (28.2 \pm 5.5)\times 10^{-12} \; \rm{erg \; cm^{-2} \; s^{-1}}$. The {\it BeppoSAX} PDS spectrum of 2001 \citep{Donato2005} is in good agreement with this high energy spectrum of {\it Swift}/BAT.

\section{Temporal study}

Regular monitoring observations with ATOM starting in 2006 
with 7 observations in 2006, 47 in 2007, 86 in 2008, 61 in 2009, and 24 in 2010
 show no significant variations ($< 1\%$) in the R band over five years of
 observations. The average brightness is 
$m_{\rm{R}} = 16.39 \pm 0.01 \; \rm{mag}$ and $m_{\rm{B}} = 18.38 \pm 0.02\; \rm{mag}$.

\begin{figure}[h]
\centering
\includegraphics[width=\columnwidth]{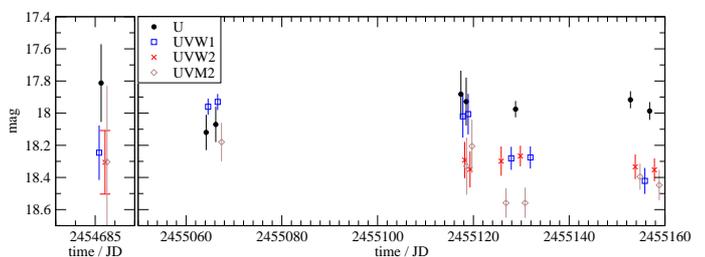}
\caption{Ultraviolet fluxes measured by {\it XMM-Newton}  and {\it Swift}  for 1ES 0229+200 in 2008 (average over 3 days) and 2009
(U: black circles, UVW1: blue open squares, UVW2: red crosses, UVM2: brown open diamonds).
The flux points at JD 2455064.6, 2455066.6, 2455117.8, and 2455118.9  are shifted by $\pm 0.4$ days for better visibility. Marginal variations are detected in the UVW1 and UVM2 bands.
} 
\label{UV_LC}
\end{figure}

The UV results from {\it XMM-Newton}  and {\it Swift}  of 2008 and 2009 are shown in Fig.~\ref{UV_LC}. 
The U and UVW2 band emission was stable and does not show significant variation ($p_{\chi^2} = 70\% $ and $p_{\chi^2} = 98\% $, respectively.).
The data in the UVW1 and UVM2 bands display a marginally significant drop by $\sim 30\%$ and $\sim 20\%$, respectively, at JD 2455120, around the time of an increase in the X-ray flux (see below).

The X-ray flux of 1ES 0229+200 instead shows an increase by a factor of around two starting at JD 2455120, as can be seen in Fig.~\ref{Xray_LC}. 
For all of 2010, only a marginal variation could be detected. During 2008-2010, marginally significant spectral changes could be detected as variations in the photon index ($p_{\chi^2} = 1\% $ for a constant) that were uncorrelated with the flux variation (see Fig.~\ref{Xray_LC}). During all epochs, the source can be well-described with a power-law model and an absorption in excess of the Galactic absorption, as described in section 3. 
At energies above 14keV, measured by {\it Swift}/BAT, marginally significant variation is detected in the monthly lightcurve with a probability for the fit of a constant of $p_{\chi^2} \approx 1\% $ over the 58 months of observation\footnote{available on http://heasarc.gsfc.nasa.gov/docs/Swift/results/bs58mon/} (\cite{Baumgartner2010}).
However, since a long integration time is needed, a behaviour similar to that in the 2-10 keV range would be difficult to detect.

\section{Spectral energy distribution}

\begin{figure}[h]
\centering
\includegraphics[width=\columnwidth]{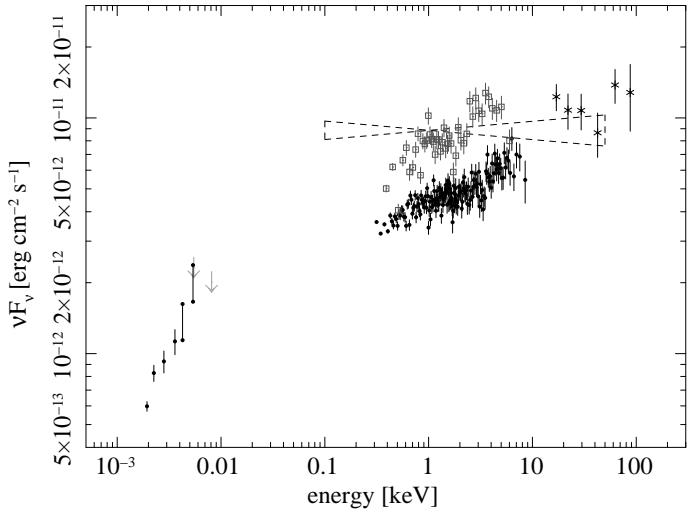}
\caption{Synchrotron emission of 1ES 0229+200 with simultaneously measured optical, UV, and X-ray emission by ATOM and {\it XMM-Newton} of August 21, 2009 (black dots). The optical and UV emission is corrected for both the host galaxy and Galactic extinction, and the X-ray emission observed by {\it XMM-Newton}/MOS is corrected for the absorption.
The bars in the UVW1 and UVM2 bands are explained in section 5.1.
The grey open squares represent the non-simultaneous {\it Swift}/XRT spectrum (corrected for detected absorption) with the highest flux in 2009. The {\it Swift}/BAT 58 month (Dec. 2004 - Oct. 2009) spectrum is shown in the energy band $>10$ keV (black crosses). The dashed line (butterfly) represents the {\it BeppoSAX} spectrum from 2001 \citep{Donato2005}.
The grey upper limits represent historical UV observations  with {\it GALEX} that are extinction corrected and of origin discussed in the text.
} 
\label{SED_Sy}
\end{figure}

\begin{figure}[h]
\centering
\includegraphics[width=\columnwidth]{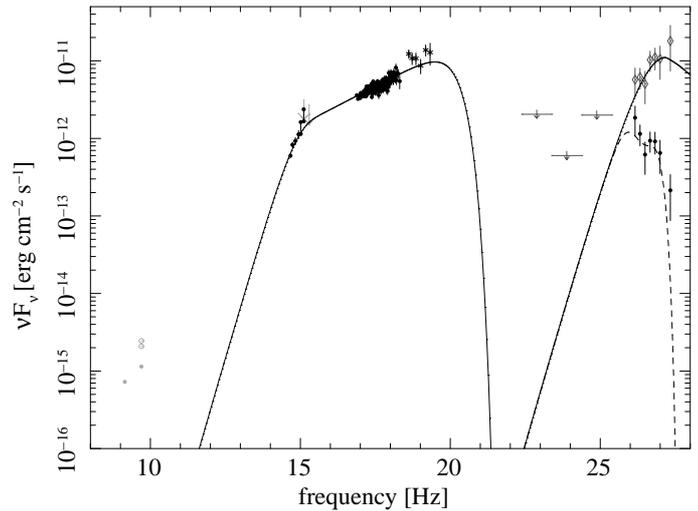}
\caption{Spectral energy distribution of 1ES 0229+200 with simultaneous measured optical, UV, and X-ray fluxes, all corrected for host galaxy emission, Galactic extinction, and Galactic absorption is shown as black data points. 
The 58 months {\it Swift}/BAT spectrum is shown $>10$ keV (black crosses). 
In grey (filled and open circles), historical radio and UV data are shown and their origin is discussed in the text.
The grey upper limits in the GeV energy range are taken from \cite{Dermer2011} and represent the upper limit in the energy bins $0.1-1\;\rm{GeV}$,  $1-10\;\rm{GeV}$, and  $10-100\;\rm{GeV}$ by {\it Fermi} observations (Aug. 2008 to Sep. 2010). 
The VHE $\gamma$-ray spectrum measured by H.E.S.S. (black circles, taken from \cite{Aharonian2007}), as well as the EBL corrected, intrinsic source spectrum (grey open diamonds) with a hard photon index, which implies an inverse Compton emission peaking at very high frequency ($>10^{27}\;\rm{Hz}$), is shown.
The solid line represents an SSC model that can describe the intrinsic synchrotron and inverse Compton emission of 1ES 0229+200, and the dashed line represents the absorption by the EBL (details about these models are described in the text).} 
\label{SED_all}
\end{figure}

The simultaneous observations in the optical by ATOM, the UV, and X-rays by {\it XMM-Newton}  are considered to study the synchrotron spectrum (see Fig.~\ref{SED_Sy}, simultaneous data of August, 21 2009). 

\subsection{Host galaxy of 1ES 0229+200}

The host galaxy of 1ES 0229+200 is an elliptical galaxy with a brightness of $m_{\rm{host,R}}  = 15.85 \pm 0.01\;\rm{mag}$  and a half-light radius of $r_e = 3.25\pm 0.07''$ 
 \citep{Urry2000}. Other observations in the Bessel R-band with the Nordic Optical Telescope (NOT) \citep{Falomo1999} show results with  $m_{\rm{host,R}} = 15.76 \; \rm{mag}$ and  $r_{\rm{e,R}} = 4  ''$.
The host galaxy profile of 1ES 0229+200 was also studied in the Bessel U, B, and V-bands with the Nordic Optical Telescope (NOT) by \cite{Hyvoenen2007}. The results are $m_{\rm{host,B}} = 18.75\;\rm{mag}$, $m_{\rm{host,U}} = 18.83\;\rm{mag}$, and $m_{\rm{host,V}} = 17.58\;\rm{mag}$ with half light radii  $r_{\rm{e,B}} =  5.65 ''$, $r_{\rm{e,U}} =  2.75''$, and $r_{\rm{e,V}} =  4.90''$.
%
%
In order to correct for the host galaxy light, a de Vaucouleurs profile of the galaxy was assumed and the measured brightness was transformed to that of an aperture of $4''$ using equation (4) of \cite{Young1976}. ATOM photometry was performed with a $4''$  aperture. 
%
%
The resulting host-galaxy corrected fluxes would result in an unphysical bump in the V band in the SED (in $\nu F_{\nu}$) since the calculated influence of the host galaxy in the V band is only $30\%$, while in the R and B band it is $90\%$ and $30\%$, respectively.
Hence, 
the influence of the host galaxy in the R, B, V, and U filters were also calculated using a spectral template for a nearby elliptical galaxy by \cite{Fukugita1995}.
Here, the R-band magnitudes and the half-light radius detected by \cite{Falomo1999} were used.
The influence of the host galaxy is then $\sim 90\%$, $\sim 80\%$, and $\sim 57\%$ for R, V, and B, respectively, which are the values that we used to correct the measured fluxes (shown in Fig.~\ref{SED_Sy}). 
The host-galaxy corrected flux in the R-band is compatible with the detected nucleus magnitude of \cite{Falomo1999}  and \cite{Urry2000} as expected from the absence of variability. 
The host galaxy influence in the UVW1 and UVM2 bands is unknown. Figure \ref{SED_Sy} therefore shows two values, connected by a bar. The upper ones correspond to the measured values
corrected only for extinction, the lower ones also assume a correction for the host galaxy of $30\%$ (the value derived in the adjacent U band).

In an independent check, the spectral slope was extracted using the Sloan Digital Sky Survey (SDSS) observations (five bands taken simultaneously). 
The resulting slope is identical to the one measured by ATOM.
Since the data do not match the epoch of the ATOM observations, the absolute fluxes were not considered.

\subsection{Galactic extinction/absorption correction}

Since the host galaxy magnitudes were not extinction corrected, the
influence of the host galaxy was first subtracted. The extinction correction was thereafter applied to the AGN light. 

The measured UV fluxes were corrected for dust absorption using E(B-V)$=$0.135 mag \citep{Schlegel1998} and the $A_\lambda / E(B-V)$ ratios given in \cite{Seaton1979}, resulting in a correction of $70\%,57\%,48\%$ for the UVM2, UVW1 and U-band, respectively. 
For the optical filters, the values for the extinction were derived from the interstellar reddening curve and table given by \cite{Zombeck1990}\footnote{online version http://ads.harvard.edu/books/hsaa/toc.html}.
The correction of $N_{\rm{H}}$ absorption was applied as described in Section 3.

\subsection{Historical multi-wavelength data}

{\it GALEX} observed 1ES 0229+200 at October 29, 2007 in the far UV ($152.8 \;\rm{nm}$) and near UV ($227.1 \;\rm{nm}$) leading to a measured flux of $31.68  \;\rm{\mu Jy}$ and $50.36  \;\rm{\mu Jy}$, respectively, taken from GalexView\footnote{http://galex.stsci.edu/GR6}.
The measured near and far UV fluxes were corrected for dust absorption using E(B-V)$=$0.135 mag \citep{Schlegel1998} and the $A_\lambda / E(B-V)$ ratios given in \cite{Seaton1979}. A correction of $64\%$ and $68\%$ resulted for the far and near UV bands, respectively.
Owing to a lack of information about the host galaxy influence in these wavebands (this influence should be very small compared to the extinction correction), the measured fluxes are shown as upper limits for the intrinsic synchrotron spectrum of 1ES 0229+200 (shown as grey arrows in Fig.~\ref{SED_Sy}).

In 2006, 2008, and 2009 {\it Integral} also observed 1ES 0229+200 several times for approximately $66 \;\rm{ks}$ with ISGRI ($17-80 \;\rm{keV}$) and $61\;\rm{ks}$ with JEM-X ($3-10 \;\rm{keV}$), but the source appears faint and too few photons were detected so that no reasonable light curve or spectrum could be extracted\footnote{see e.g. http://www.isdc.unige.ch/heavens\_webapp/integral}.

In historical snapshot observations of VLA at 6 cm in 1992 fluxes of $41.5 \;\rm{mJy}$ \citep{Schachter1993} and $49.09 \;\rm{mJy}$ \citep{Perlman1996} were detected (shown as grey open circles in Fig.~\ref{SED_all}). With 6 cm VLBA observations this flux was resolved into a core of $22.7 \;\rm{mJy}$ and a jet of $7.7 \;\rm{mJy}$ \citep{Rector2003}. In the same study, the core flux at 1.4 GHz was detected by VLA observations to be $51.8 \;\rm{mJy}$, and 1ES 0229+200 was found to have curved jets to the north and south with extensions of $30''$. 
The core fluxes are shown as grey filled circles in Fig.~\ref{SED_all}.

\subsection{SSC model}

\cite{Katarzynski2006} demonstrated that synchrotron spectra with a high
minimum Lorentz factor can explain a very hard inverse Compton spectra.
The simultaneously obtained data in the optical, UV, and X-ray bands cover the whole synchrotron emission.
This allows an empirical determination of the minimum Lorentz factor in 1ES~0229+200. 
The high accuracy of the {\it XMM-Newton} measurements constrains the synchrotron spectrum very
efficiently.

The photon index $\Gamma = \alpha +1 $ of the X-ray spectrum gives a direct estimate of the spectral index $n$ of the electron distribution $N(E)dE \propto E^{-n}dE$ with the relation $n=-1-2\alpha$. A broken power law with a canonical cooling break of $\Delta n = 1$ is assumed for the electron distribution. 
The variability of 1ES~0229+200 detected in the X-ray emission places constraints on the maximum size of the emission volume considered in the SSC model which is inferred to be 
 $R< (\Delta t/\rm{sec})\times D \times c \approx D \times 2.6\times 10^{14} \;\rm{m}$.

The peaks in the SED are commonly explained by
leptonic models, such as those of \cite{Marscher1985}, as synchrotron and inverse Compton
(IC) emission from a population of relativistic electrons up-scattering their
self-produced synchrotron photons (synchrotron self-Compton models (SSC)). The
code of \cite{Krawczynski2004} for a one-zone SSC model was used to
describe the intrinsic emission of 1ES 0229+200. For this model, a spherical
emission volume of radius R, moving with a bulk Doppler factor D towards the
observer and a magnetic field B was assumed. The electron distribution was
described by a broken power-law between minimum and maximum energy.

The X-ray spectrum can be fitted well by a single power-law model over 1.5
orders of energy. The extrapolation fits the absorption and host-galaxy
corrected UV data, suggesting that there is a single power law slope for the energy range $0.005 -100\;\rm{keV}$. The absorption and host galaxy corrected UV-optical emission below this range is significantly steeper and strongly constrains the minimum
energy of the electron distribution function,
yielding a very high value.
The well-determined spectral index in the X-ray regime has to describe the low-energy spectral index of the electron distribution $n_1=-2.6$.
The uncooled electron spectral slope hence has $n_2 =-3.6$ following the canonical break.
The radius was kept as the maximum value obtained from the variability timescale ($R=1\times 10^{16}\;\rm{m}$).
This radius is an upper limit and is fixed independently of the SED
  modelling. Smaller values of radii would also be consistent with the
  variability constraint. \cite{Tavecchio2009} used a smaller value in
  their attempts to describe the SED compiled in their
  study. The SED shown in Fig. 5 cannot be reproduced with a radius ($R <
  1\times 10^{15}\;\rm{m}$) (see below).
To account for the high energy peak of the IC emission, the underlying electron distribution must be very narrow. 
Together these constraints are best met with the model parameters
$E_{\rm{min}} = 2\times 10^{11}\;\rm{eV}$ ($\gamma_{\rm{min}}=3.9\times 10^5$), $E_{\rm{break}} = 3.2\times 10^{13}\;\rm{eV}$ ($\gamma_{\rm{break}}=6.2\times 10^7$), $E_{\rm{max}} = 1\times 10^{14}\;\rm{eV}$ ($\gamma_{\rm{max}}=1.9\times 10^8$), and a magnetic field of  $B=3.2\times10^{-5} \;\rm{G}$.
The Doppler factor of D=40 was chosen as the smallest one possible to reproduce the spectra.
This model can describe the simultaneously obtained spectra in the synchrotron regime, the long-term hard X-ray spectrum by {\it Swift}/BAT, and the non-simultaneous, intrinsic TeV spectrum.
The peak frequency of the synchrotron and IC emission based on the described model are 
$\nu_{\rm{sy,peak}} = 3.5 \times 10^{19} \;\rm{Hz}$ and 
$\nu_{\rm{IC,peak}} = 1.5 \times 10^{27} \;\rm{Hz}$. 
The peak fluxes show a slight Compton dominance with 
$\nu_{\rm{sy,peak}} F_{\nu_{\rm{sy,peak}}} = 9.7 \times 10^{-12} \; \rm{erg \; cm^{-2}\;s^{-1}}$ and 
$\nu_{\rm{IC,peak}} F_{\nu_{\rm{IC,peak}}} = 1.1 \times 10^{-11} \; \rm{erg \; cm^{-2}\;s^{-1}}$. 
%
This model is shown in Fig.~\ref{SED_all} as a solid line, while the dashed line represents the absorption by the EBL using the model of \cite{Franceschini2008}.
This model is in good agreement with the EBL absorption model used in \cite{Aharonian2007} to correct the observed TeV spectrum (shown in Fig.~\ref{SED_all}).
%

The cooling break is at very high energies such that a single power-law model for the electron distribution with $n=-2.6$ fits the data equally well  
and is an alternative model to describe the measured spectra of the SED.

An earlier attempt to utilize the suggestion of \cite{Katarzynski2006} was reported by  \cite{Tavecchio2009}. They assumed a  narrow single power-law electron distribution with a high minimum Lorentz factor to fit the data. 
As in the modelling presented here, the slope of the synchrotron emission
below $\gamma_{\rm{min}}$ is given by $F_{\rm{sy}} \sim \nu^{1/3}$, leading to a slope for the IC part that is similar to that is required to fit the TeV spectrum.
Attempting to fit the overestimated X-ray flux,  \cite{Tavecchio2009}
  derived a different set of parameters, including a higher (factor 10)
  magnetic field and a smaller radius. The initial model presented above used
  an upper limit to the radius, which was not optimized in the SED fitting.
In an attempt to explore models with smaller radii, attempts were made to
reproduce the SED with smaller radii. The  precise determination of the low
energy cutoff in the UV range and the well-determined spectral shape in the
X-ray range can only be reproduced by invoking a very high Doppler factor
(D$>100$). While providing an adequate fit to the synchrotron range, they
do not reproduce the inverse Compton emission peak well.

The cut-off of the synchrotron emission above 100 keV represents the maximum energy of the electron distribution. No change in spectral shape could be detected in the {\it Swift}/BAT spectrum.
Therefore only a lower limit to the maximum cut-off energy could be determined
empirically. The precise measurement of the maximum energy would require a
soft $\gamma$-ray detection. In the IC part of the spectrum, the maximum energy cannot be
determined emprically either because any increase in $E_{\rm{max}}$ does not result 
 in a change in the IC spectrum because of Klein-Nishina suppression.
Assuming that the high energy cut-off of the synchrotron emission is at the
high energy end of the {\it Swift}/BAT spectrum ($\sim 100\;\rm{keV}$), this
results in an electron distribution that extends only over five orders of
energy.

Changes in the peak fluxes of the synchrotron emission (such as the difference between the XMM-Newton and the {\it Swift}/XRT spectrum, as shown in Fig.~\ref{SED_Sy}) were also tested.
Any changes in the parameters of the emission volume corresponding to an increase in
flux by a factor of $\sim 2$ yield a change in the inverse Compton peak flux
of a factor of $\sim 2$ as well. This implies that changes in the VHE fluxes
in the TeV energy range are also expected to occur in 1ES 0229+200.

Alternative models were also explored. Apart from the attempts
to describe the SED by SSC emission from a very compact source, other
assumption would also lead to models involving a high Doppler factor.
Assuming that the derived curvature in the optical regime would be represented by the break of the electron distribution with 
$n_1=-1.6$, $n_2=-2.6$, a separate model can be constructed. This scenario
cannot be excluded by our simultaneously measured SED above
$10^{14}\;\rm{Hz}$,
but would require a large Doppler factor of about D=100 in order to account 
for the high energy peak of the IC emission.
%
This alternative model also underpredicts the radio flux of the resolved core
by a large factor, e.g. around $70\%$.
Interestingly, the radio flux of the jet, as measured by \cite{Rector2003}, could be represented by such model, indicating that the whole synchrotron emission originates from the resolved jet.
Another possible way of explaining the high IC peak using such an electron distribution function, but with a lower Doppler factor, would be to invoke an additional external Compton contribution resulting from the scattering  of the CMB as discussed in \cite{Boettcher2008} for  1ES~1101-232, a distant TeV blazar with a hard TeV spectrum. This model would decouple the X-ray and TeV emission because most of the TeV emission would result from the interaction with the CMB further away from the synchrotron emission region of the jet. 
This model would imply no VHE variability on short timescales. 
The effect of the CMB scattering depends on the distance to the source, so that it is more likely to be detectable in distant sources such as 1ES~0229+200 and 1ES~1101-232. 


\section{Conclusions}

The X-ray spectral information that have been presented here cover all available X-ray data up to October 2010. 
Together with the simultaneous UV and optical observations, they yield a very good coverage of the synchrotron emission.
Our detailed study of the high quality {\it XMM-Newton}  spectrum has inferred an
absorption higher than the nominal Galactic column density, which could be intrinsic to the source or caused by Galactic excess absorption.

The host-galaxy, extinction-corrected optical and UV fluxes have been shown to provide strong evidence that the cut-off of the low energy part of the synchrotron emission is located between the optical and X-ray regime. Therefore, the minimum energy of the electron distribution has to be rather high to account for this cut-off. 
As suggested by \cite{Katarzynski2006}, an electron distribution with a high minimum Lorentz factor is needed to reproduce a hard TeV spectrum.
A narrow electron distribution, as indicated by the high minimum energy,
results in a hard intrinsic VHE $\gamma$-ray spectrum as deduced for 1ES~0229+200. This hard VHE spectrum also shows that the inverse Compton peak is at very high energies ($>10^{27}\;\rm{Hz}$), hence the initial electrons are very energetic.

An earlier attempt to utilize the suggestion by \cite{Katarzynski2006} was reported by  \cite{Tavecchio2009}, assuming a high minimum Lorentz factor. 
The main differences between the model presented here and their attempts
  are caused by the new, precisely determined low-energy cut-off in the optical 
regime and the simultaneously determined X-ray spectrum.

A broken power-law could be an alternative description to the additional absorption found in the X-ray spectra. However, the low energy extrapolation of the X-ray spectrum would not fit the optical range and is therefore disfavoured.

The hard X-ray spectrum up to 15 keV together with the long-term spectrum by {\it Swift}/BAT have been found to show that the synchrotron peak is extended up to $\sim 100 \; \rm{keV}$ without any significant cut-off in the X-ray spectrum.

1ES 0229+200 is defined as a high-frequency peaked BL Lac object, and the measured synchrotron emission peaks at higher frequencies ($>$ 100 keV) than usual for HBL and belongs therefore to the class of extreme blazars. The exact peak frequency cannot be determined, since the hard X-ray spectrum does not show any significant cut-off. 
Assuming that the high energy cut-off of the synchrotron emission is at the high energy end of the {\it Swift}/BAT spectrum ($\sim 100\;\rm{keV}$), the underlying electron distribution extend only over 
five orders of magnitude in energy, which is a narrow range. For no other extreme blazar has this range been demonstrated to be so narrow.

The low energy part of the synchrotron emission has not been found to show significant
variation over four years in the optical R band.
Only minor variation has been detected in the ultraviolet flux during 2008 and 2009. The 2-10 keV X-ray flux 
instead varied by a factor of $\approx 2$ within around 20 days in 2009. In the high energy part of the synchrotron emission, measured by {\it Swift}/BAT, variations in this amplitude and timescale could not be detected because of the longer integration time needed. 



Several blazars are known as so-called extreme blazars with synchrotron emission extending to high energies \citep{Costamante2001}. 
It has been found that  Mkn 501 revealed a very high energy peak of synchrotron emission around 100~keV in a flare with a detected large shift in the peak frequency compared to previous observations \citep{Pian1998}. The monitoring by  {\it BeppoSAX} of 1ES 2344+514 has identified huge changes in the synchrotron peak frequency within one year with different spectral shapes \citep{Giommi2000}. {\it INTEGRAL} observations of 1ES 1426+428 display a synchrotron peak frequency around 100~keV \citep{Wolter2008}. These extreme blazars have similar synchrotron peak frequencies to 1ES 0229+200, 
but their very high energy spectra in the TeV range are much softer than for 1ES 0229+200.
Hence, 1ES 0229+200 has the highest IC peak frequency among the extreme blazars known up to date.



\begin{acknowledgements}
The authors acknowledge the support by the {\it XMM-Newton} team in arranging simultaneous observations.
The execution and availability of the {\it RXTE}  and {\it Swift}  observations and the use of the public HEASARC software packages are acknowledged. 
S.K. and S.W. acknowledge support from the BMBF through grant DLR 50OR0906.
\end{acknowledgements}
%
\bibliographystyle{bibtex/aa}
\bibliography{ref_0229.bib}



\section{Supplement}
For the online supplement:

\begin{table*}
\centering
\begin{tabular}{|c|c|c|c|}
\hline
Name & Coordinates (J2000) & $F_{\mbox{0.3-12 keV}} \;(\rm{erg \; cm^{-2}\; s^{-1}})$& Possible counterpart (radio, IR, optical)\\
\hline
1ES 0229+200 & 02:32:48.592,+20:17:16.15 & $(3.71 \pm 0.04) \times 10^{-11}$  & NVSS 023248+201716, GSC2.2 N3312301936 \\
XMMU 023228.6+202349 & 02:32:28.690,+20:23:49.10 & $(2.0 \pm 0.2) \times 10^{-12}$  & GSC2.2 N33123012274 \\
XMMU 023227.3+200711 & 02:32:27.371,+20:07:11.61 & $(1.2 \pm 0.2) \times 10^{-12}$  & NVSS 023227+200711, GSC2.2 N331230310473 \\
XMMU 023147.2+201348 & 02:31:47.287,+20:13:48.93 & $(8 \pm 2) \times 10^{-13}$  & GSC2.3 NC6R014254 \\
XMMU 023318.0+201237 & 02:33:18.050,+20:12:37.19 & $(5 \pm 3) \times 10^{-13}$  & -- \\
XMMU 023212.7+200318 & 02:32:12.738,+20:03:18.55 & $(4 \pm 2) \times 10^{-13}$  & -- \\
XMMU 023315.5+201323 & 02:33:15.554,+20:13:23.39 & $(4 \pm 2) \times 10^{-13}$  & NVSS 023314+201330, IRAS 02304+2000 \\
XMMU 023229.9+200639 & 02:32:29.933,+20:06:39.66 & $(3 \pm 1) \times 10^{-13}$  & GSC2.2 N331230310392 \\
XMMU 023314.7+200626 & 02:33:14.715,+20:06:26.90 & $(2 \pm 1) \times 10^{-13}$  & -- \\
XMMU 023248.2+202858 & 02:32:48.277,+20:28:58.01 & $(2 \pm 1) \times 10^{-13}$  & -- \\
XMMU 023315.2+202553 & 02:33:15.208,+20:25:53.65 & $(2.1 \pm 0.9) \times 10^{-13}$  & GSC2.3 NC6P013245 \\
XMMU 023230.4+201704 & 02:32:30.426,+20:17:04.60 & $(2.1 \pm 0.5) \times 10^{-13}$  & -- \\
XMMU 023233.2+200539 & 02:32:33.215,+20:05:39.75 & $(2 \pm 1) \times 10^{-13}$  & GSC2.3 NC6R010173 \\
XMMU 023219.6+201201 & 02:32:19.684,+20:12:01.90 & $(1.6 \pm 0.9) \times 10^{-13}$  & NVSS 023219+201204 \\
XMMU 023323.5+202140 & 02:33:23.568,+20:21:40.04 & $(1.6 \pm 0.8) \times 10^{-13}$  & GSC2.3 NC6P013116 \\
XMMU 023321.4+202330 & 02:33:21.475,+20:23:30.11 & $(1.2 \pm 0.6) \times 10^{-13}$  & GSC2.3 NC6P002203 \\
XMMU 023314.7+200852 & 02:33:14.791,+20:08:52.69 & $(0.3 \pm 1) \times 10^{-13}$  & GSC2.3 NC6R010904 \\
XMMU 023254.6+202352 & 02:32:54.603,+20:23:52.74 & $(9 \pm 5) \times 10^{-14}$  & -- \\
XMMU 023307.2+201040 & 02:33:07.248,+20:10:40.15 & $(8 \pm 4) \times 10^{-14}$  & GSC2.3 NC6R011298 \\
XMMU 023328.0+202342 & 02:33:28.066,+20:23:42.91 & $(8 \pm 6) \times 10^{-14}$  & IRAS 02306+2010 \\
XMMU 023310.8+201929 & 02:33:10.871,+20:19:29.55 & $(5 \pm 3) \times 10^{-14}$  & GSC2.3 NC6P001373 \\
XMMU 023211.7+201721 & 02:32:11.773,+20:17:21.75 & $(1 \pm 3) \times 10^{-14}$  & GSC2.2 N3312301924 \\
\hline
\end{tabular}
\caption{X-ray sources detected in the {\it XMM-Newton} observation on 1ES 0229+200. Their possible counterparts in the radio, IR, and optical regimes were searched in the Guide Star Catalog (GSC) \citep{Lasker2008}, the IRAS faint source catalog \citep{Moshir1990}, and the NRAO VLA Sky Survey (NVSS) \citep{Condon1998}.
}
\label{tab_srcdet}
\end{table*}

\end{document}